\documentclass[a4paper,UKenglish,cleveref, autoref, thm-restate]{lipics-v2021}

\usepackage{booktabs}
\usepackage{multirow}
\usepackage{tabularx}
\usepackage{tcolorbox}
\usepackage{hyperref}

\title{Operationalizing the EU AI Act in Agile Software Development: A Guideline-Based Approach} 

\titlerunning{Operationalizing the EU AI Act in Agile Software Development} 

\author{Dennis Schrader}{University of Applied Sciences and Arts Hannover, Ricklinger Stadtweg 120, 30459 Hannover, Germany}{schraderde@gmail.de}{https://orcid.org/0009-0003-8432-9865}{}
\author{Eva-Maria Schön}{University of Applied Sciences Emden/Leer, Constantiaplatz 4, 26723 Emden, Germany}{eva-maria.schoen@hs-emden-leer.de}{https://orcid.org/0000-0002-0410-9308}{}
\author{Henning Fritzemeier}{Volkswagen AG, Wolfsburg, Germany}{henning.fritzemeier@volkswagen.de}{}{}
\author{Michael Neumann}{University of Applied Sciences and Arts Hannover, Ricklinger Stadtweg 120, 30459 Hannover, Germany}{michael.neumann@hs-hannover.de}{https://orcid.org/0000-0002-4220-9641}{}

\authorrunning{D. Schrader et al.}



\nolinenumbers 

\EventEditors{John Q. Open and Joan R. Access}
\EventNoEds{2}
\EventLongTitle{42nd Conference on Very Important Topics (CVIT 2016)}
\EventShortTitle{CVIT 2016}
\EventAcronym{CVIT}
\EventYear{2016}
\EventDate{December 24--27, 2016}
\EventLocation{Little Whinging, United Kingdom}
\EventLogo{}
\SeriesVolume{42}
\ArticleNo{23}

\begin{document}
\newcolumntype{L}{>{\raggedright\arraybackslash}X}
\maketitle

\begin{abstract}
\textit{Context:} The EU AI Act requires providers and deployers of Artificial Intelligence (AI) systems to implement documentation, risk management, and human oversight. Agile teams that ship AI features in short iterations lack specific artifacts to discharge these duties, since the regulation's abstract provisions do not map onto the Definition of Done, Sprint Reviews, or working agreements. 
\textit{Objective:} We provide agile teams with an actionable compliance instrument: an evaluated guideline that operationalizes EU AI Act obligations as activities integrable into existing agile practice. We further document the translation method behind it so that the approach can be reused for adjacent regulations.
\textit{Method:} Following Design Science Research, we assessed each EU AI Act article along three dimensions (technical, normative, organizational). We subsequently classified the articles using a traffic-light scheme and mapped those deemed highly relevant to previously documented pain points of agile teams working with AI. We validated the resulting catalog with practitioners through a survey and 11 additional semi-structured expert interviews, analyzed via qualitative content analysis.
\textit{Results:} The guideline comprises 12 items covering roles and responsibilities, risk and quality management, transparency and traceability, monitoring, and regulatory sandboxes. Practitioners rated the catalog as understandable and relevant; feasibility varied with organizational maturity. Effective adoption towards EU AI Act compliance requires collective ownership across roles and integration into existing agile events rather than parallel compliance processes.
\textit{Conclusions:} The catalog gives agile teams a starting point to transform their delivery practices towards an EU AI Act compliance without dismantling agile practices. Our underlying translation method is reusable for regulations such as DORA, NIS2, and GDPR. A replication package is publicly available.
\end{abstract}

\section{Introduction}
Organizations developing or integrating Artificial Intelligence (AI) components in agile or hybrid software development face a compliance problem that is now operational rather than prospective. The EU AI Act~\cite{EUAIAct2024} entered into force in August~2024, and the core obligations for high-risk AI systems apply from 2~August~2026~\cite[Art.~113]{EUAIAct2024}. Non-compliance carries administrative fines of up to 35\,M\,EUR or 7\,\% of global annual turnover, whichever is higher~\cite[Art.~99]{EUAIAct2024}. Providers and deployers in the EU market therefore have to discharge lifecycle-wide duties for risk management, technical documentation, data governance, transparency, and human oversight~\cite[Art.~9,~11,~13,~14]{EUAIAct2024} on a fixed clock, while delivery teams continue to ship in two- or three-week sprints.

The friction is not at the management layer, where frameworks such as the NIST AI Risk Management Framework~\cite{NISTAIRMF} and ISO/IEC~42001~\cite{ISO42001} provide organizational controls. It
is at the team layer, where the regulation's obligations must turn into specific entries. For instance, in a specific aspect of the Definition of Done, adoption in a Sprint Review, items in backlog refinement, or rows in a Working Agreement. The EU AI Act is process-neutral by design: it prescribes outcomes, not ceremonies. A team applying  a software process (e.g., Scrum, Kanban, or SAFe) receives no guidance on whether the \textit{Art.~9} risk register lives in the backlog or beside it, whether \textit{Art.~13} transparency obligations expand the user story template, or whether \textit{Art.~14} human oversight is the Product Owner's, the team's, or a separate role's responsibility. Recent empirical work confirms that this gap is observable in practice: Neumann et al.~\cite{NeumannEtAl2026XP} show in their study how regulatory pressures translate into policies that practitioners then bypass to maintain agile velocity, and Sillberg et al.~\cite{SillbergEtAl2025} synthesize practitioner sentiment in a multivocal review and report widespread uncertainty about team-level implementation.

Similar research has produced compliance artifacts at the product, audit, or organizational level, which we discuss in Section~\ref{sec:related}. None of these contributions operates on
the team layer, where the obligations meet the cadence of agile delivery, and the one approach that addresses an agile artifact is scoped to a single organization and lacks broader validation. What is missing is a team-level instrument that translates the obligations onto the agile events and artifacts an arbitrary team already uses, together with empirical evidence that practitioners
find such an instrument understandable, relevant, and feasible to adopt. 

Three research questions guide this work:

\noindent\textbf{RQ1:} \textit{How can the requirements of the EU AI Act be systematically translated into actionable guidelines for the use of AI in agile software development?}\\
Our first research questions aim for the translation between abstract regulatory text and team-level practice. We answer it by creating a reproducible procedure that assesses each EU AI
Act article along three dimensions, classifies relevance through a traffic-light scheme, and maps high-relevance articles onto documented agile pain points~\cite{BahiEtAl2024,Herda.2025}. The expected outcome is both a method
that is reusable for adjacent regulations and a concrete guideline catalog derived from it.\\
\noindent\textbf{RQ2:} \textit{How do practitioners assess the guidelines regarding understandability, relevance, and feasibility?}\\
RQ2 verifies whether the artifact resulting from RQ1 holds up under practitioner scrutiny on three constructs that are diagnostic for an applied instrument: a) whether the guideline can be read without further interpretation, b) whether it captures the obligations that matter for agile teams, and c) whether it can be enacted in the carrier artifacts those teams already maintain. We expect this to surface design weaknesses and to indicate which clusters are robust across organizational contexts.\\
\noindent\textbf{RQ3:} \textit{Which organizational factors influence the adoption of our EU AI Act adoption guideline in agile teams?}\\
The last research question RQ3 moves from artifact quality to the conditions under which adoption succeeds or fails. We probe how organizational size, regulatory exposure, agile maturity, and existing artifacts (Definition of Done, Working Agreements, recurring events) shape the path from acknowledging a guideline to embedding it. The expected outcome is a set of adoption conditions that practitioners and researchers can use to anticipate where the guideline can be integrated readily and where preparatory work will be required.

The paper makes three core contributions: First, we propose an approach for translating regulatory articles into specific (agile) practices. The procedure assesses each article along a technical, a normative, and an organizational dimension, classifies its relevance for agile teams through a traffic-light classification, maps known challenges while adopting or using agile methods from scientific literature, and aggregates them into thematic clusters. The approach can be adapted to regulations with comparable structure such as DORA, NIS2, and the GDPR; we sketch this transfer in Section~\ref{sec:discussion}. Second, we present a EU AI Act adoption guideline consists of 12~items organized into five clusters: roles and responsibilities, risk and quality management, transparency and traceability, monitoring and continuous compliance, and regulatory sandboxes. Third, we evaluated our guideline based on a survey and semi-structured interviews with 11 practitioners to verify its applicability in practice.

The paper at hand is structured as follows: Section~\ref{sec:related} positions our work against AI governance frameworks and related studies. Section~\ref{sec:design} details the translation procedure and the validation design. In Section~\ref{sec:guideline}, we present the guideline catalog and in Section~\ref{sec:results} the validation results. Section~\ref{sec:discussion} discusses implications for practice, lessons learned, and the transfer of the methodology to other regulations. Section~\ref{sec:ttv} addresses threats to validity before the paper closes with a conclusion and future work in Section~\ref{sec:conclusion}.

\section{Related Work}
\label{sec:related}
We position our work along three lines of research: the use of AI in agile software development and the operationalization of the EU AI Act for software engineering (SE). We close by stating the gap that our guidelines address.

\subsection{AI in Agile Software Development}
\label{sec:related:ai-agile}
Empirical studies report that generative AI (GenAI) changes day-to-day agile work. Ulfsnes et al.~\cite{UlfsnesEtAl2024} observe shifts in collaboration, pair work, and individual workflows once AI assistants enter the sprint. Bahi et al.~\cite{BahiEtAl2024} review agile project management pain points that AI tooling may mitigate, from estimation support to backlog refinement. Russo et al.~\cite{RussoEtAl2024CopenhagenManifesto} argue in the Copenhagen Manifesto that GenAI in software engineering must remain human-centered, naming oversight, accountability, and developer agency as design constraints. Herda et al.~\cite{Herda.2025} summarize the XP 2025 workshop in which practitioners and researchers map AI-and-agile challenges to a research roadmap.

Three recent contributions sharpen this picture. Plan\"otscher et al.~\cite{PlanoetscherEtAl2026XP} survey 73 agile practitioners across seven software activities and report that AI is currently used as an assistant or not used at all, with practitioners expecting a shift toward AI as a collaborator. Perkusich et al.~\cite{PerkusichEtAl2026XP} survey 70 Brazilian practitioners on LLM use in Scrum management and find broad daily use alongside
concerns about ``almost correct'' outputs, confidentiality, and hallucinations. Neumann et al.~\cite{NeumannEtAl2026XP} conduct a multiple case study in three German organizations and document how regulatory pressures translate into policies that practitioners then bypass to maintain agile velocity; one of their cases classifies GenAI tools by the EU AI Act risk pyramid, which is the most direct empirical evidence to date that the act is already shaping agile working practice. These studies establish that AI is now an agile concern and that the regulatory dimension is perceptible to practitioners. They do not, however, prescribe how teams should encode the resulting obligations in their daily practice.

\subsection{Operationalizing the EU AI Act for Software Engineering}
\label{sec:related:euaiact-se}
Several recent papers translate the EU AI Act into software engineering artifacts. Wagner and
colleagues~\cite{WagnerBorgRuneson2024IEEESW,WagnerEtAl2025REW,WagnerEtAl2024PROFES} introduce the regulation to the IEEE Software readership, dissect its requirements from an SE perspective, and add an industrial case study on high-risk readiness. Kelly et al.~\cite{KellyEtAl2024IEEECAI} 
extend the ISO/IEC 25059 product quality model with attributes derived from the act for safety-critical AI. Hupont et al.~\cite{HupontEtAl2024UseCaseCards} contribute an UML-based use case
card to document the intended purpose of a system in line with the risk classification, while M\"okander et al.~\cite{MoekanderEtAl2022ConformityAssessments} discuss conformity assessment and post-market monitoring as auditing instruments. H\"uger et al.~\cite{HuegerEtAl2023LoD} report from Cariad on an instantiation of the act through a Level of Done layer for ML projects in the automotive domain. Sovrano et al.~\cite{SovranoEtAl2025} study how LLM-based tooling can support the drafting of technical documentation required by the act. Sillberg et al.~\cite{SillbergEtAl2025} synthesize practitioner and academic sentiment toward the act in a multivocal review, and Westerstrand~\cite{Westerstrand2025Fairness} argues that compliance does not exhaust ethical obligation and gives recommendations for fairness in agile AI development. Tuape et al.~\cite{TuapeEtAl2025CONISOFT} address Article \textit{14 head-on and design mechanisms for human oversight and accountability} in AI-driven software engineering. Genovesi~\cite{Genovesi2025AIGovernance} describes the introduction of an AI governance framework in a financial organization as a practitioner report.

Each of these contributions advances a specific compliance artifact (a product quality model, a documentation template, an audit procedure, a Level of Detail (LoD) layer, a governance framework, a SE perspective on the regulation). None of them targets the team-level agile working practices through which providers and deployers actually meet their day-to-day obligations.


\subsection{Research Gap}
\label{sec:related:gap}
To the best of our knowledge, no published academic work translates the EU AI Act obligations onto the agile practices and artifacts an arbitrary team uses, or evaluates such a translation with practitioners across organizational contexts. 

To be more precise, we identified three main gaps: First, the AI-and-agile literature recognizes that AI changes agile work and that the regulatory dimension is already perceptible in practice, with our own prior case study~\cite{NeumannEtAl2026XP} documenting how teams improvise around the absence of clear rules; none of this work, however, supplies a team-level instrument that encodes the act's obligations. Second, the work on EU AI Act operationalization produces compliance artifacts at product, audit, or organizational level, and the one approach that explicitly addresses an agile artifact, H\"uger et al.'s Level of Done~\cite{HuegerEtAl2023LoD}, remains scoped to automotive ML and is not empirically validated with practitioners beyond the case organization. Third, agile-in-regulated-environments studies demonstrate feasibility for other regulations but precede the EU AI Act. Our paper addresses these gaps.

\section{Research Design}
\label{sec:design}
\begin{figure}[h!]
    \centering
    \includegraphics[width=1\linewidth]{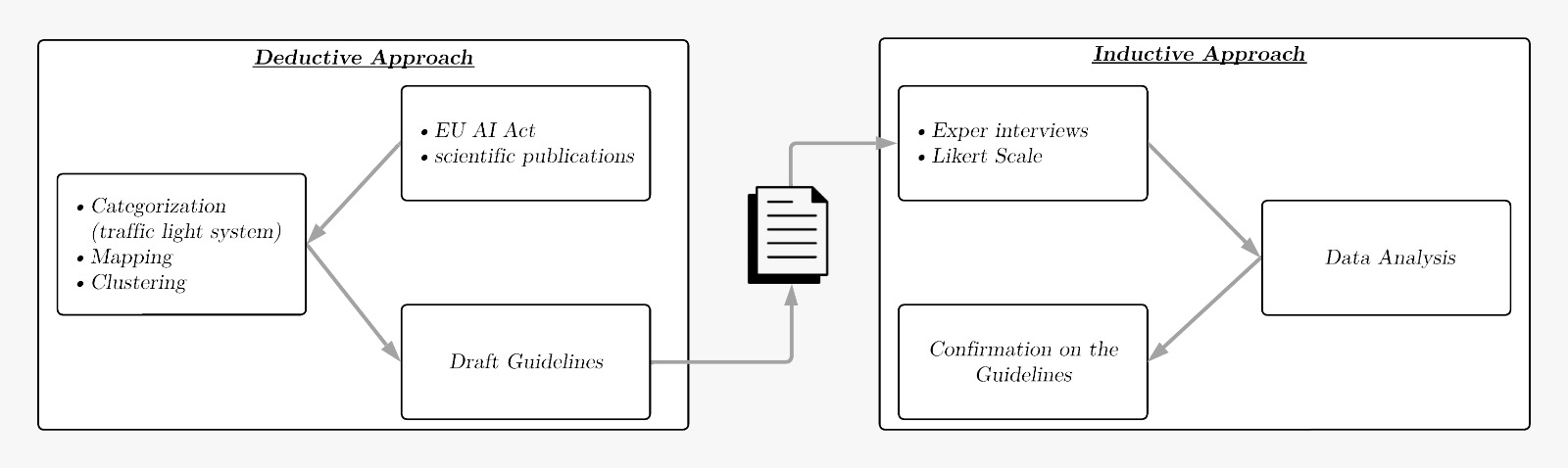}
    \caption{Research Design for the development and evaluation of our guidelines}
    \label{fig:ResearchDesign}
\end{figure}

We designed our study by applying Design Science Research (DSR)~\cite{Hevner2004,Wieringa.2014,Engstroem.2020} and organize it along the three cycles of Hevner~\cite{Hevner2004}. DSR fits this work because the contribution is an artifact, a translation procedure and the resulting guideline, constructed and then validated with its target community, which is the configuration DSR was developed for in software engineering~\cite{Engstroem.2020,Wieringa.2014}. The \emph{relevance cycle} is grounded in a concrete industry problem: agile teams covered by the EU AI Act must satisfy obligations from
02.08.2026 onward~\cite[Art.~113]{EUAIAct2024} without abandoning iterative delivery. The \emph{design cycle} produces a reproducible translation procedure and a guideline of 12 items
(Section~\ref{sec:guideline}). The \emph{rigor cycle} validates both with a Likert survey and eleven semi-structured expert interviews analyzed through qualitative content analysis~\cite{Mayring2022}. Figure~\ref{fig:ResearchDesign} depicts our research design.

\subsection{Artifact Creation}\label{sec:methods}
The creation of the guideline proceeded in four phases (A--D). Each phase delivered an
intermediate artifact for the next: a coded article set, a traffic-light classification, a mapping table, and the cluster structure.

\textbf{Phase A: Deductive Analysis of the EU AI Act:} We analyzed all articles of Regulation (EU)~2024/1689 along three dimensions: 
\begin{itemize}
    \item a \emph{technical} dimension (data, monitoring, tooling),
    \item a \emph{normative} dimension (obligations attached to provider or deployer roles under Art.~16 and~26),
    \item an \emph{organizational} dimension (roles, governance, process integration).
\end{itemize}

Each dimension is anchored in recent empirical work on AI in agile contexts: technical
in AI-assisted defect detection and test automation~\cite{Todorov2022}, normative in the alignment between AI Act obligations and ISO/IEC standards~\cite{KellyEtAl2024IEEECAI}, organizational in roles, governance, and oversight in agile AI delivery~\cite{BahiEtAl2024,Herda.2025}. Coding criteria were derived deductively from the research questions and these references, following category-system construction as in Bortz and D\"oring~\cite{BortzDoering2006} and the DSR guidance of
Engstr\"om et al.~\cite{Engstroem.2020}.

\textbf{Phase B: Traffic-light Classification:} For each article and each dimension, we assigned \emph{high}, \emph{medium}, or \emph{low} relevance. \emph{High} means the article imposes obligations a team must address inside its development workflow; \emph{medium} means the article is contextually relevant but generates no direct team-level work; \emph{low} means it concerns
governance levels beyond the team or actors other than providers or deployers. The aggregated rating per article is the maximum of the three dimensional ratings, since a single high-relevance dimension is sufficient to make an article actionable. Articles with ambiguous coding were re-examined and, where uncertainty persisted, assigned the medium label so that ambiguity does not silently propagate.

\textbf{Phase C: Mapping to Agile Pain Points:} We mapped every article rated high in at least one dimension to documented agile pain points of agile teams working with AI~\cite{BahiEtAl2024,Herda.2025}, consolidated from the four studies cited above into recurring themes such as unclear AI ownership, lack of structured human oversight, insufficient data governance, and compliance documentation accumulating outside agile artifacts. The mapping is many-to-many and was refined iteratively to avoid uncovered pain points and redundant article coverage. We provide an example in Section~\ref{sec:guideline}.

\textbf{Phase D: Cluster Building and Guideline Derivation:} We grouped the mapped articles into thematic clusters following phases~1--5 of the type-building qualitative content analysis of
Kuckartz and R\"adiker~\cite{KuckartzRaediker2024}. Clustering produced five clusters (\emph{C1 Roles and Responsibilities}, \emph{C2 Risk and Quality Management}, \emph{C3 Transparency and Traceability}, \emph{C4 Monitoring and Continuous Compliance}, \emph{C5 Regulatory
Sandboxes and Innovation} under Art.~57). Within each cluster, we derived items at a granularity that lets a team address them inside an existing event or artifact (sprint planning, refinement,
review, retrospective, Definition of Done, Working Agreements); this rule was applied iteratively until 12 items remained.

\subsection{Practitioner Validation}\label{sec:methods:validation}
Validation followed a sequential mixed-methods design: a Likert survey first, expert interviews second, so the interviews could probe the survey ratings. Both data collection instruments are available in the research protocol which we made available~\cite{Schrader.2026a}.

\textbf{Likert Survey:} A 5-point Likert survey on LimeSurvey rated each of the 12 survey items
on three constructs: \emph{understandability}, \emph{relevance} for agile software development, and \emph{feasibility} in agile teams (1 = strongly disagree, 5 = strongly agree). Each cluster was
introduced with a short context block paraphrasing the EU AI Act obligations it addresses, so participants rated items in context rather than in isolation.

\textbf{Expert Interviews:} We conducted eleven semi-structured expert interviews online via
Microsoft Teams between 20 October 2025 and 25 November 2025. Participants were recruited through the professional networks of the third and fourth author of this paper and one public IT practitioner forum to diversify the sample beyond a single organization, with diversity sought along role, industry, and organization size. Sampling proceeded in two waves. The first wave covered six participants (P1--P6); we monitored category
emergence after each interview and observed no new categories after the fifth interview, which we treat as theoretical saturation in the sense of Mayring~\cite{Mayring2022}. We then conducted five
additional interviews (P7--P11) to confirm saturation and broaden contextual variation across roles, industries, and organization sizes; no new main categories emerged in this second wave.
Table~\ref{tab:sample} provides an overview of the interviewee profiles and their characteristics.

\begin{table}[t]
  \centering
  \caption{Interview sample (N=11). Pseudonyms used throughout the
  paper. Organization sizes are banded for anonymization.}
  \label{tab:sample}
  \footnotesize
  \begin{tabularx}{\linewidth}{@{}l L l l c c@{}}
    \toprule
    ID  & Role                                       & Domain                              & Org.\ size & AI exp. & AI Act \\
        &                                            &                                     & (empl.)    & (yrs)   & fam.   \\
    \midrule
    P1  & Lead Agile Coach / RTE                     & Automotive                          & $>$100k    & 2--3 & low--med    \\
    P2  & Sales \& Key Account Manager               & IT Services / Custom Dev.           & 1k--5k     & 3--5 & high        \\
    P3  & Software Developer                         & IT / Software Development           & n/a        & 2--4 & low         \\
    P4  & AI \& Innovation Manager                   & Public Administration               & $>$10k     & 1--2 & high        \\
    P5  & Agility Master                             & Railway / Critical Infrastructure   & $>$100k    & 1--3 & medium      \\
    P6  & Former Head of IT / IT Consultant          & Manufacturing / GRC                 & 0.5k--1k   & 3--5 & low--med    \\
    P7  & Lead Agile Coach                           & Affiliate Marketing / SE            & 1k--5k     & 2--4 & med--high   \\
    P8  & Head of Transformation                     & Insurance                           & 0.5k--5k   & 1--2 & low--med    \\
    P9  & CEO / Innovation Consultant                & Innovation Consulting / SW Dev.     & 10--50     & 3--5 & medium      \\
    P10 & Engineering Manager                        & Affiliate Marketing / SE            & 1k--5k     & 2--3 & medium      \\
    P11 & Head of Digital Solutions                  & Industrial / E-Commerce / SAP       & $>$1k      & 1--3 & medium      \\
    \bottomrule
  \end{tabularx}

  \smallskip
  \footnotesize
  Org.\ size is banded into discrete ranges (10--50, 0.5k--1k, 1k--5k,
  0.5k--5k, $>$1k, $>$10k, $>$100k employees) to preserve anonymity
  while retaining contextual variation. AI Act familiarity is
  self-reported on a five-step scale: low, low--medium, medium,
  medium--high, high.
\end{table}

The interview guide had four blocks. Blocks~1 (background and AI exposure), 2 (general assessment), and~4 (closing feedback) were identical across interviews. Block~3 was individualized per interview from the participant's prior survey responses (the \emph{expectation profile}) to probe divergences between low and high ratings on the same item or cluster. Interviews were held online using Microsoft Teams, were recorded with consent, transcribed automatically with \href{https://voice.ai/}{Voice AI}, and anonymized before analysis. The shortest interview had a duration of 18:35 minutes, whereas the longest interview lasted 43:11 minutes. The average interview duration was 30:00 minutes; more precisely, the mean duration was 29:59.8 minutes. This variation reflected the survey-based data, specifically the expectation profiles, on which the interviews were based. Interviews were generally shorter when there was broad overall agreement, whereas greater discrepancies required more extensive discussion. The interview guide template and the per-interview instantiations of block~3 are in the replication package.

\subsection{Qualitative Data Analysis}\label{sec:methods:qca}
We analyzed the transcripts with structuring qualitative content analysis following Mayring~\cite{Mayring2022}. A first pass per interview produced superordinate codes; a second pass consolidated them into main and subcategories across interviews; a third pass applied the resulting coding guide (definitions, coding rules, anchor examples) to the full material. The coding guide is also available in the research protocol~\cite{Schrader.2026a}.

Coding was performed by the first author. We did not compute inter-coder reliability; the implication is discussed in Section~\ref{sec:ttv}. As a partial compensation, the
interviewer applied in-situ communicative validation by summarizing key statements back to participants and asking for confirmation. Likert data were analyzed descriptively (means, medians, full distributions); inferential statistics are omitted because $n=11$ does not support them.

\section{The Guideline}
\label{sec:guideline}
This section presents the artifact produced by the design cycle and answers RQ1: \textit{How can the requirements of the EU AI Act be systematically translated into actionable guidelines for the use of AI in agile software development?} We report the outcome of the traffic-light classification, the five clusters, the 12 items that constitute the guideline, and the mapping to agile elements\footnote{Here we reference an Agile Element as an agile practice or artifact in accordance with Neumann~\cite{Neumann.2021}.}. The guideline~\cite{Schrader.2026} is the input to the validation reported in Section~\ref{sec:results}. 

\subsection{Outcome of the Traffic-Light Classification}
\label{sec:guideline:traffic}
We assessed all 113 articles of the EU AI Act along the three dimensions (technical, normative, and organizational) introduced in Section~\ref{sec:methods}. Articles classified as \emph{high relevance} on at least one dimension entered the mapping step; articles rated \emph{medium} or \emph{low} on all three were retained for
traceability but did not generate items. The full classification is part of the replication package (see Section~\ref{sec:availability}).

Three observations shape the guideline. First, high-relevance articles cluster in the high-risk obligations of Chapter~III (Art.~9--15) and the transparency provisions of Chapter~IV (Art.~50), with concentration on the normative and organizational dimensions. Second, the technical dimension
attracts fewer high-relevance ratings than the other two; technical obligations are largely indirect, mediated by documentation and process requirements. Third, \textit{Art.~57 regulatory sandboxes} is the only article that addresses a voluntary mechanism rather than an obligation, which justifies treating it as a separate cluster (Section~\ref{sec:guideline:clusters}).


\subsection{Five Clusters Across the AI Adoption Regulation}
\label{sec:guideline:clusters}
The mapping step (Phase~C, Section~\ref{sec:methods}) linked each high-relevance article to one or more agile pain points reported by Bahi et al.~\cite{BahiEtAl2024} and Herda et al.~\cite{Herda.2025}. The cluster step (Phase~D) aggregated articles and pain points along thematic proximity, following the type-building qualitative content analysis according to Kuckartz and R\"adiker~\cite{KuckartzRaediker2024}, Phases~1--5. The result is five clusters that span the guideline.

\textbf{C1 Roles \& Responsibilities} translates the provider and deployer obligations of Art.~3, 16, and 26 into team-level role definitions. Agile teams already operate with defined roles (Product Owner, Scrum Master, developers); the cluster does not introduce new ones but binds existing roles to AI-related accountabilities.

\textbf{C2 Risk \& Quality Management} covers \textit{Art.~9 risk management}, \textit{Art.~10 data governance}, and \textit{Art.~14 human oversight}. The cluster extends iterative risk identification into AI-specific concerns and embeds output review into existing reflection points.

\textbf{C3 Transparency \& Traceability} addresses \textit{Art.~11 and~13 technical documentation and instructions of use} and the transparency obligations of Art.~50. This cluster addresses the tension between the agile preference for minimal upfront documentation and the EU AI Act
transparency duties, making AI influence on work products visible without re-introducing heavyweight specification.

\textbf{C4 Monitoring \& Continuous Compliance} responds to \textit{Art.~17 quality management} and \textit{Art.~72 post-market monitoring}. The cluster treats compliance as a continuous activity rather than a stage-gate.

\textbf{C5 Regulatory Sandboxes \& Innovation} addresses \textit{Art.~57 regulatory sandbox}. It differs from C1--C4 because sandboxes are voluntary. The cluster positions sandboxes as a low-threshold entry point for AI experimentation under regulatory observation.

\subsection{The Twelve Items of the AI Adoption Guideline}
\label{sec:guideline:items}
Table~\ref{tab:guidelines} lists the 12 items, the cluster each belongs to, and the EU AI Act articles they operationalize. We use the form ``\emph{teams + active verb + agile element + traceable outcome}'' to keep each item actionable. Each item is short enough to
fit a Definition of Done entry or a Working Agreement line.

\begin{table}[h]
\centering
\caption{Items of the EU AI Act-aligned guideline for agile teams.
Article numbers refer to Regulation (EU) 2024/1689.}
\label{tab:guidelines}
\small
\begin{tabularx}{\linewidth}{@{}llcX@{}}
\toprule
\textbf{ID} & \textbf{Cluster} & \textbf{Art.} & \textbf{Item} \\
\midrule
G1.1 & C1 Roles \& Resp.       & 3, 16, 26 & Responsibilities for AI use are assigned explicitly and aligned with EU AI Act provider and deployer roles. \\
G1.2 & C1 Roles \& Resp.       & 3, 16, 26 & Role assignments are documented in agile artifacts such as Working Agreements or the Definition of Done. \\
G2.1 & C2 Risk \& Quality      & 10        & Data quality is checked and documented continuously within the team. \\
G2.2 & C2 Risk \& Quality      & 9         & AI-related risks are identified, assessed, and documented iteratively. \\
G2.3 & C2 Risk \& Quality      & 14        & Outputs of AI systems are reviewed regularly under human oversight. \\
G3.1 & C3 Transparency         & 11, 13    & Work products are documented such that the influence of AI is traceable. \\
G3.2 & C3 Transparency         & 50        & Transparency about AI influence is established in recurring agile events (e.g., Sprint Review). \\
G4.1 & C4 Monitoring           & 17, 72    & Retrospectives are used to reflect on regulatory changes and integrate them into the team's way of working. \\
G4.2 & C4 Monitoring           & 17, 72    & Teams establish a procedure to verify and document compliance continuously. \\
G5.1 & C5 Sandboxes            & 57        & AI systems are first deployed in sandbox environments before productive use. \\
G5.2 & C5 Sandboxes            & 57        & Sandbox results are reflected in agile events such as Sprint Reviews or retrospectives. \\
G5.3 & C5 Sandboxes            & 57        & Sandbox use is documented to satisfy traceability and accountability requirements. \\
\bottomrule
\end{tabularx}
\end{table}

The guideline addresses obligations on providers and deployers symmetrically. Where the EU AI Act differentiates the two roles, the items remain neutral and leave role attribution to G1.1 in Table~\ref{tab:guidelines}.

\subsection{Agile Elements Mapping}
\label{sec:guideline:carriers}
The guideline does not prescribe new agile practices. We map each item to one or more existing agile element (Table~\ref{tab:carriers}) so that adoption requires adapting current artifacts rather than introducing parallel structures. This mapping is the design decision that operationalizes the agile pain point of process overhead outlined in the related work (see Section~\ref{sec:related}).

\begin{table}[htbp]
\centering
\caption{Mapping of items to agile practices and artifacts.}
\label{tab:carriers}
\small
\begin{tabularx}{\linewidth}{@{}lX@{}}
\toprule
\textbf{Agile Element} & \textbf{Items} \\
\midrule
Working Agreement       & G1.1, G1.2 \\
Definition of Done      & G1.2, G2.1, G3.1, G4.2 \\
Backlog Refinement      & G2.1, G2.2 \\
Sprint Review           & G2.3, G3.2, G5.2 \\
Retrospective           & G2.2, G4.1, G4.2, G5.2 \\
Decision log / ADR      & G3.1, G5.3 \\
Sandbox protocol        & G5.1, G5.2, G5.3 \\
\bottomrule
\end{tabularx}
\end{table}

The mapping is not bijective. Most items attach to more than one carrier, and most carriers carry more than one item. We treat this as a feature rather than a limitation: it preserves the flexibility that agile teams expect. Taken together, the traffic-light classification, the five clusters, the 12 items, and the carrier mapping form the artifact answering RQ1.

\section{Learnings from the Empirical Validation}
\label{sec:results}
As stated in Section~\ref{sec:design}, we validated the guideline through a Likert-based survey and eleven semi-structured expert interviews. Section~\ref{sec:results-validation:assessment} reports the practitioner assessment along the three construct dimensions and answers RQ2. Section~\ref{sec:results-validation:adoption} reports the organizational factors that emerged from the interview analysis and answers RQ3.

\subsection{Practitioner Assessment of Understandability, Relevance, and Feasibility}
\label{sec:results-validation:assessment}
The second research question, RQ2, asks how practitioners assess the guideline along the three construct dimensions understandability, relevance, and feasibility: \textit{How do practitioners assess the guidelines regarding understandability, relevance, and feasibility?} Across the eleven participants the guideline is verified understandable and relevant; assessments of feasibility are more dispersed and depend on the team's working context rather than on the wording of individual items. We base this assessment on the Likert ratings shown in Table~\ref{tab:likert} and on the follow-up interviews that probed deviations from the expectation profile.

\textbf{Understandability} is the dimension on which the guideline is least contested. P10 framed the contrast that recurs in most interviews: condensed items are easier to act on than the full regulatory text. Only two of the 12 items attract any disagreement at all, and the dissent concentrates on terminology rather than substance, G1.2 references the \emph{Working Agreement}, an artifact that several participants reported as unfamiliar in their own teams. P11 made the same point in a different register, asking whether the documentation duty was tied to a specific carrier or could be discharged equivalently elsewhere. The interviews position the guideline as a translation device: practitioners read it not as a new vocabulary but as an explicit version of practice that was already implicit.

\textbf{Relevance} disperses more than understandability and tracks the mandatory, voluntary distinction in the underlying articles. The items that operationalize Chapter~III obligations of the EU AI Act, especially \textit{human oversight} (G2.3) and \textit{iterative risk identification} (G2.2), are rated as clearly relevant and prompted no contrary statements in the interviews. By contrast, the \textit{continuous-compliance} item G4.1 and the \textit{sandbox-documentation} item G5.3 attract the largest share of neutral or disagreeing responses. The qualitative material identifies two distinct mechanisms behind this. For G4.1 the criticism is one of locus: P11 rejects the retrospective as the venue for regulatory discussion, arguing that such topics ``require organizational coordination rather than team-internal self-regulation''; P10 expresses the same reservation with regard to mandating regulatory reflection inside an event that is already crowded. For G5.3 the criticism is one of scope: sandbox documentation is seen as relevant where teams build AI systems but not where they merely use AI tools, a distinction (``developing AI'' versus ``developing with AI'') that P10 and P11 both raise independently.

\textbf{Feasibility} is the dimension that exposes the guideline most sharply to the working context. The single strongest objection in the entire validation, the only \emph{strongly disagree} rating, is on
G3.1 (documenting the influence of AI on work products), placed by P2 and accompanied by three further low ratings. The reservation is not about the principle, as transparency regarding AI influence is endorsed across the sample, but about the practical effort required in teams that have no established mechanism for documenting it. P3 captured the boundary condition that runs through the feasibility judgments: the items are realistic
only when they ``become part of daily work and are not perceived as an additional burden.'' Where existing agile carriers (Definition of Done, Sprint Review, Retrospective) are active, integration is described as incremental; where they are weak or absent, even the most basic documentation item becomes contentious.

We summarize the distributional view and report medians and interquartile ranges in Table~\ref{tab:likert}. A consistent pattern runs through the three dimensions. The guideline holds up where it formalizes practices that agile teams recognize (risk identification, human oversight, peer review) and meets resistance where it adds documentation or process duties that fall on the team itself (G3.1, G4.1) or addresses voluntary mechanisms (G5.3). TThe median rating is 4 for almost all items: dimension combinations, indicating broad agreement. The main empirical signal therefore lies in the variation of responses, which shows where practitioners' assessments diverged.

In summary, practitioners assess the guideline as understandable and broadly relevant, with feasibility contingent on the team's agile carriers.


\begin{table}[t]
\centering
\caption{Likert ratings per item ($N = 11$). $\textit{Mdn}$ = median, $\textit{IQR}$ = interquartile range. Five-point scale (1 = strongly disagree, 5 = strongly agree).}
\label{tab:likert}
\small
\setlength{\tabcolsep}{4pt}
\begin{tabular}{@{}lccccccc@{}}
\toprule
& \multicolumn{2}{c}{\textbf{Understandability}}
& \multicolumn{2}{c}{\textbf{Relevance}}
& \multicolumn{2}{c}{\textbf{Feasibility}} \\
\cmidrule(lr){2-3} \cmidrule(lr){4-5} \cmidrule(lr){6-7}
\textbf{ID} & $\textit{Mdn}$ & $\textit{IQR}$
            & $\textit{Mdn}$ & $\textit{IQR}$
            & $\textit{Mdn}$ & $\textit{IQR}$ \\
\midrule
G1.1 & 4 & 1.0 & 4 & 1.0 & 4 & 0.0 \\
G1.2 & 4 & 1.5 & 4 & 1.0 & 4 & 0.5 \\
G2.1 & 4 & 1.0 & 4 & 0.5 & 4 & 0.0 \\
G2.2 & 5 & 0.5 & 4 & 1.0 & 4 & 1.0 \\
G2.3 & 5 & 0.0 & 5 & 1.0 & 5 & 1.0 \\
G3.1 & 4 & 1.0 & 4 & 1.0 & 3 & 1.0 \\
G3.2 & 5 & 1.0 & 4 & 1.0 & 4 & 1.0 \\
G4.1 & 4 & 1.0 & 3 & 1.5 & 3 & 1.0 \\
G4.2 & 5 & 1.0 & 4 & 1.0 & 4 & 1.0 \\
G5.1 & 4 & 1.5 & 4 & 1.5 & 4 & 1.5 \\
G5.2 & 4 & 1.0 & 4 & 1.0 & 4 & 0.0 \\
G5.3 & 4 & 1.0 & 3 & 0.5 & 4 & 0.0 \\
\bottomrule
\end{tabular}
\end{table}

\subsection{Organizational Factors Shaping Adoption}
\label{sec:results-validation:adoption}
Here, we answer RQ3: \textit{Which organizational factors influence the adoption of the guidelines in agile teams?} Based on our results, we identified six categories that describe the conditions under which the guideline is taken up or rejected. In the brackets we provide the information how often we identified the aspects to the category in the interviews.

\textbf{Function and benefit of the guideline (10/11).} The guideline is positioned as an instrument of orientation rather than as a checklist. P1 described it as ``a good basis to engage with AI more systematically and consciously''; P9 framed the underlying regulation as ``an opportunity for quality and clarity, not as a brake.'' The shared interpretation from the participants is that the guideline does not require entirely new ways of working but makes existing agile practices more explicit and links them to EU AI Act compliance. Adoption is reported to be easier where teams already accept this framing and harder where the guideline is read as an additional rulebook.

\textbf{Implementation and integration (8/11).} Whether the guideline takes hold depends on whether it is embedded into the artifacts that already structure the team's work. For example, \emph{Working Agreement}, \emph{Definition of Done}, \emph{Sprint Review}, and \emph{Retrospective}. P3 articulated the operational boundary condition explicitly: the items work only when they ``become part of daily work and are not perceived as an additional burden.'' P10 reported that quality assurance for AI-generated code
already runs through existing reviews, and that the value of the guideline lies in making this explicit rather than in adding ceremonies. P8 named the converse failure mode: mere availability is
not enough, and introduction needs change management and active support. Tooling and training were named as enablers across the sample; absent or implicit ownership of compliance was the most common blocker.

\textbf{Roles, responsibility, and communication (8/11).} The guideline is read as a redistribution of accountability, not as the assignment of compliance to a single role. Participants rejected the model of
delegating AI compliance to the Product Owner, the Scrum Master, or a dedicated compliance officer in isolation. P4 emphasized that the guideline helps ``make responsibilities visible, especially at the
leadership level''; P9 added that formal role documentation must not replace individual ownership: ``regulatory accountability cannot substitute for personal responsibility.'' Two participants proposed
concrete role designs: P10 suggested AI champions to coordinate compliance topics; P11 framed a defined AI responsibility as ``a major opportunity for the company.'' Across the sample, communication between development teams and compliance functions was named as the mechanism that converts shared responsibility into observable practice.

\textbf{Agility versus regulation (7/11).} Participants did not treat the guideline as a threat to agile values and  principles, but they articulated the latent risk. P2 warned that strict rules ``slow down the agile process''; P11 added that documentation duties ``can slow down development processes.'' The risk is reduced when the items extend existing events and amplified when compliance is migrated into
parallel meetings. Participants endorsed integration into the Sprint Review and the Retrospective as a way to keep compliance recurrent without breaking cadence, with one notable dissent: P11 explicitly
rejected the Retrospective as the venue for regulatory discussion, arguing that such topics belong to organizational coordination, not team self-regulation. This dissent maps onto the lowest relevance
rating in the guideline (G4.1) and is the clearest qualitative explanation for that quantitative outlier.

\textbf{Context and organizational maturity (7/11).} A one-size-fits-all interpretation did not hold in practice. The same item was perceived as straightforward in one organization and difficult to implement in another, depending on organizational maturity, resource availability, and the strength of existing agile artifacts. P1 contrasted large organizations, where ``the items work well,'' with smaller teams that ``need more individual approaches''; P9 reported the opposite from a different perspective, arguing that smaller teams may adopt the items more bindingly because they are closer to the actual work. P8 noted that the guideline ``does not directly fit a non-agile context'' and therefore requires translation. P10 and P11 independently raised a distinction that the guideline itself does not explicitly make: \emph{developing AI} and \emph{developing with AI} are different cases, and some clusters, notably C5 on sandboxes, are more relevant to the former than to the latter. According to the participants, industry was less important than organizational maturity: heavily regulated industries did not automatically find the guideline easier to adopt; what mattered was whether the regulatory function and the development teams already had an established mode of collaboration.

\textbf{Participation and continuous evolution (5/11).} The guideline is read as a living instrument, not as a static document. P1 summarized that competencies ``cannot be conveyed once but must be built up continuously''; P4 added that practitioners themselves should be involved in revising the guideline. P7 and P10 emphasized reflection in agile events as the operational mechanism; P8 named explicit change management, training sessions, and consultation hours. Static publication of the guideline, without ownership for its ongoing revision, was identified as a likely failure mode.

\paragraph*{Cross-category patterns.}
Three patterns cut across the six categories. First, organizational maturity, operationalized as the presence and active use of agile artifacts, is the strongest covariate of perceived feasibility with the AI Adoption Guideline. Teams with active artifacts describe integration as incremental; teams without them describe it as a precondition that has to be established first. Second, industry regulation builds awareness but not capacity: participants in regulated industries did not report greater ease in adopting the AI Adoption Guideline. Third, organization size shapes which carriers participants regard as realistic: smaller organizations rely on the Working Agreement as a single agile carrier, larger organizations distribute the items across more specialized artifacts. With $N = 11$ these patterns are working hypotheses for further studies, not statistical claims. Taken together, three factors are decisive for the adoption of the AI Adoption Guideline: collective ownership of compliance rather than delegation to a single role, embedding into existing agile events, and the maturity of the agile team's artifacts.


\section{Discussion \& Implications for Practice}
\label{sec:discussion}

\begin{tcolorbox}[
  colback=gray!8,
  colframe=black!70,
  boxrule=0.5pt,
  arc=2pt,
  left=8pt, right=8pt, top=6pt, bottom=6pt,
  title=\textbf{Key Findings},
  fonttitle=\bfseries
]
Guidelines bridge regulation and agile delivery only when agile teams embed them in existing artifacts and events rather than running a parallel governance track. Effectiveness depends on organizational maturity and on responsibility for AI compliance being shared across the team rather than delegated to a single role. A fixed set of guidelines does not scale to organizational diversity; practitioners need a stable backbone together with explicit guidance for context-specific adaptation.
\end{tcolorbox}

\subsection{Interpretation of Findings}
\label{subsec:interpretation}
We interpret the interview results along three patterns that emerged consistently across the sample.

First, practitioners read the guidelines as a translation layer rather than as new policy. The function described most often across the sample is making existing practice explicit and reducing uncertainty about what Articles~9, 11, 14, and 17 of the EU AI Act~\cite{EUAIAct2024} require inside an iteration and the applied agile practices in it. This reading is consistent with Article~96 EU AI Act~\cite{EUAIAct2024}, which positions guidelines as instruments of uniform application rather than as substantive obligations.

Second, adoption depends on contextual fit. The implementation and integration category was raised by 8 of 11 participants and is the strongest predictor of perceived feasibility. Teams with established agile artifacts described integration as incremental, whereas teams without such artifacts need to create or formalize suitable structures before the AI Adoption Guideline can be embedded in day-to-day work. Industry regulation built awareness but not capacity: participants in regulated industries did not report easier adoption. P10 and P11 surfaced a distinction the guidelines themselves do not make explicit, between \emph{developing AI} and \emph{developing with AI}. Cluster~C5
(Regulatory Sandboxes) is more pertinent to the former, while C3 (Transparency) and C4 (Monitoring) carry weight for both.

Third, responsibility for AI compliance is collective in practice but unassigned by default (8 of 11). Participants rejected delegation of EU AI Act obligations to the Product Owner, the Scrum Master, or a dedicated compliance officer in isolation. P4 stressed that the guidelines help make responsibilities
visible at the leadership level; P9 added that formal role documentation cannot substitute for individual ownership. Two participants proposed concrete designs: P10 suggested AI champions to coordinate compliance topics, P11 framed defined AI responsibility as ``a major opportunity for the company.''
The absence of a default owner is a known agile pain point~\cite{BahiEtAl2024}; cluster~C1 (Roles and Responsibilities) encodes collective ownership in role extensions and in agile artifacts rather than in a new compliance role.

The pattern on agility versus regulation refines this picture. The risk of compliance dragging on delivery decreases when the guidelines extend existing events and increases when compliance moves into parallel meetings. One dissent on this point is informative rather than incidental: P11 explicitly
rejected the Retrospective as the venue for regulatory discussion, arguing that such topics belong to organizational coordination, not team self-regulation. This single position is the clearest qualitative explanation for the lowest relevance rating in the Likert results (G4.1) and marks the limit of automatic integration into agile events.

\subsection{Positioning our Contribution to the Related Work}
\label{subsec:positioning}
Our results complement existing work along four areas. Todorov~\cite{Todorov2022} maps AI techniques onto software engineering tasks but does not address regulatory obligations; we add the compliance layer. Bahi et al.~\cite{BahiEtAl2024} identify agile pain points around AI adoption without linking them to legal requirements; we close that link through the traffic-light mapping between high-relevance EU AI Act articles and named pain points. Kelly et al.~\cite{KellyEtAl2024IEEECAI} translate EU AI Act requirements into technical specifications for safety-critical AI; we cover the organizational dimension their technical framing does not reach. Herda et al.~\cite{Herda.2025} report practitioner concerns from the XP\,2025 workshop but stop at the level of observations; we convert comparable concerns into a validated set of guidelines.

Two contributions are very close to our work. H\"{u}ger et al.~\cite{HuegerEtAl2023LoD} introduce a Level of Done layer for ML projects at Cariad as the most direct prior attempt to encode EU AI Act obligations in an agile artifact; their approach targets automotive ML and is not validated with practitioners outside the case organization. Neumann et al.~\cite{NeumannEtAl2026XP} document how teams in three German organizations improvise around the absence of clear rules and bypass emerging policies to maintain agile velocity. The present paper responds to that finding: the guidelines provide the team-level instrument those teams were missing.

The convergent point across this body of work is that EU AI Act compliance in agile contexts is an organizational problem, not a technical one. None of the prior contributions provides a validated set of guidelines that practitioners can apply directly to their Working Agreements and Definition of Done. The
artifact contribution of our paper addresses this gap.

\subsection{Implications for Practice}
\label{subsec:implications}
We derive seven implications. Each is grounded in interview evidence and tied to a specific thematic cluster.

\textbf{L1: Extend existing agile practices before adding new ones.} The strongest signal across the sample (8 of 11) is resistance to additional meetings or roles. Sprint Reviews, Retrospectives, and Backlog Refinements address the EU AI Act obligations of Articles~9 and~17 when scoped to include AI risk and compliance reflection. P3 stated the operational condition plainly: the guidelines work only when they ``become part of daily work and are not perceived as an additional burden.'' Compliance migrated into parallel ceremonies was reported as the first ceremony to be dropped under delivery pressure.

\textbf{L2: Anchor responsibilities in artifacts, not in job titles.} Cluster~C1 (Roles and Responsibilities) is most effective when accountability is encoded in Working Agreements and the Definition of Done rather than assigned to a named role. P9 phrased it plainly: regulatory accountability cannot substitute for personal responsibility. Artifact-level anchoring protects the team against turnover and against the false reassurance of appointing a compliance officer with no execution mandate inside the sprint.

\textbf{L3: Calibrate guideline depth to organizational maturity (7 of 11).} The same guideline produces different outcomes in teams with and without established agile practices. Organizations at lower maturity should adopt clusters~C1 (Roles and Responsibilities) and~C2 (Risk and Quality) first and defer C4 (Monitoring) and C5 (Regulatory Sandboxes) until the basic governance loop is reliable. Adopting the full set at once was reported as overwhelming. P8 named the converse failure mode for any maturity level: availability is not enough, and introduction requires change management and active support.

\textbf{L4: Make AI influence visible at the artifact level.} Cluster~C3 (Transparency) can be implemented with low overhead through short provenance markers on user stories, pull requests, and review notes. P10 reported that quality assurance for AI-generated code already runs through existing reviews and that the value of the guideline lies in making this explicit rather than in adding ceremonies. Heavyweight documentation templates were rated as the least feasible part of the guidelines.

\textbf{L5: Validate the guideline with legal and compliance functions before adoption.} The AI Adoption Guideline is intended as a team-level translation instrument, not as a substitute for legal interpretation. Before applying the items in a specific organization, teams should review the wording and operational implications together with legal, compliance, or regulatory experts. This is particularly important because industry-specific obligations, organizational policies, and legal interpretations may require adaptations to the guideline. Such review helps ensure that the simplified wording of the items does not omit legally relevant distinctions or create unintended bias in how EU AI Act obligations are interpreted.

\textbf{L6: Treat the guidelines as a living artifact (5 of 11).} Participants read the guidelines as evolving rather than static. P1 noted that competencies cannot be conveyed once but have to be built up continuously. Agile teams should plan a review cadence for the guideline (e.g., on a quarterly basis) in which the guidelines are revised against retrospective input and against changes to the EU AI Act implementation acts. A frozen set loses legitimacy.

\textbf{L7: Invest in AI literacy before invoking human oversight.} Article~14 EU AI Act requires human oversight, but oversight without competence is ceremonial. Participants linked feasible oversight to prior
training on model limitations, prompt design, and output assessment. Teams that adopt cluster~C2 (Risk and Quality) without a parallel literacy investment risk producing formally compliant but substantively hollow review records.



\section{Threats to Validity}\label{sec:ttv}
As with every study also ours comes with limitations. We applied the threats to validity concept~\cite{Wohlin.2012} to mitigate the limitations and identify specific measures to optimize the quality.
We discuss threats below.

\paragraph*{Construct Validity} The three assessment dimensions (technical, normative, organizational)
and the three Likert constructs (understandability, relevance, feasibility) were defined by the first author from methodological literature and from the EU AI Act text, without external concept
validation. Each Likert construct is measured by a single item per guideline, which limits construct coverage; we mitigate this by triangulating the ratings with semi-structured interviews that probe
the same constructs in open form. The term \emph{guideline} in this paper refers to a team-level translation artifact and is distinct from the Commission-issued guidelines foreseen in Art.~96 EU AI Act~\cite{EUAIAct2024}. 

Because the guideline items translate selected EU AI Act obligations into concise team-level formulations, this translation may introduce interpretive bias, for example if legally relevant terms are simplified, omitted, or interpreted differently across organizational contexts. The study did not include a formal legal validation of the items by legal experts or organizational legal departments. Accordingly, the AI Adoption Guideline should be understood as an operational support instrument for agile teams, not as legal advice or as a substitute for organization-specific legal assessment.

\paragraph*{Internal Validity} We addressed this through a documented coding guide with definitions, coding rules, and anchor examples and through a three-pass procedure (Section~\ref{sec:methods:qca}); both are part of the replication package. Communicative validation was partial: we used probing and summarizing during the interviews but did not return the final category system to participants for member checking. Block~3 of the interview guide was individualized from each participant's prior survey responses to target divergence between survey and interview, which may bias probing toward expected
explanations; Blocks~2 and~4 remained fixed across all interviews to preserve a common baseline.

\paragraph*{External Validity}
The sample of eleven participants is not representative. Recruitment proceeded through the professional networks of both supervisors and one public IT practitioner forum and targeted agile roles with an
organizational and methodological focus (e.g., Agile Coach, Agile Manager); pure developer, data-scientist, and compliance roles are absent. The EU AI Act is in its early implementation phase, with
obligations for high-risk systems applying from 2~August~2026, and the regulatory context is European; transfer to other jurisdictions or to later enforcement phases is not warranted by our data. Cluster
C5 addresses regulatory sandboxes under Art.~57, which are voluntary; feasibility ratings for C5 generalize only to organizations that actively consider sandbox participation.

\paragraph*{Conclusion Validity}
The Likert data are ordinal and the sample size precludes inferential statistics; we report medians and interquartile ranges and do not test differences across guidelines or constructs. The qualitative analysis
carries the interpretation margin inherent to structuring content analysis; the cross-case patterns in
Section~\ref{sec:results-validation:adoption} are working hypotheses, not statistical claims. We draw no causal conclusions about adoption.

\paragraph*{Reliability}
The replication package~\cite{Schrader.2026a} covers the traffic-light mapping, the Article-to-cluster-to-guideline mapping, anonymized Likert raw data, the expectation profile, the interview guide, the coding guide, and the anonymized analysis matrix. 

\section{Conclusion \& Future Work}
\label{sec:conclusion}
This paper addresses how agile teams can discharge EU AI Act obligations without disrupting iterative delivery practices. Three findings carry our contribution. First, the regulation's obligations can be translated into actionable guidelines for agile teams through a reproducible procedure that combines a three-dimensional article assessment, a traffic-light classification, and a mapping to documented agile pain points. Second, practitioners assessed the resulting 12 guidelines in five clusters as understandable and
relevant; perceived feasibility of adoption varied with organizational maturity and the strength of existing agile artifacts. Third, adoption is conditioned by three factors that are organizational rather than technical: collective ownership of compliance, integration into existing agile events, and treatment of the guideline as a living artifact rather than a static document.

Three directions extend this work. A larger and more diverse practitioner sample is needed to test whether the cross-category patterns observed in $N=11$ generalize across industries, organization sizes,
and maturity levels. A longitudinal pilot inside the industry partner organization would shift the evidence base from perceived to actual adoption, traceable in Working Agreements, the Definition of Done,
and the carrier artifacts of clusters C1--C5. The translation procedure is regulation-agnostic by construction; applying it to DORA, NIS2, and the GDPR would test that claim, isolate the EU AI Act-specific elements from the transferable backbone.

\section*{Data Availability Statement}\label{sec:availability}
The replication package for this study is publicly available~\cite{Schrader.2026a}. It includes the traffic-light classification, the article-to-cluster-to-guideline mapping, the expectation profiles, the customized interview guide based on the expectation profiles, the coding guide \& book, and the anonymized analysis matrix. The survey raw data and interview transcripts are not publicly available in order to protect participant confidentiality and to comply with data protection requirements. Access to additional anonymized materials may be provided by the corresponding author upon reasonable request, subject to ethical and legal constraints.

We also made the guideline as a companion document available~\cite{Schrader.2026}.

\section*{Acknowledgements}
We thank all participants who participated in the survey and expert interviews for their time and valuable insights.

\bibliographystyle{plainurl}
 \bibliography{references}
\end{document}